\documentclass[aps,pra,twocolumn,showpacs]{revtex4}
\usepackage{graphicx}\usepackage{amssymb}\usepackage{hyperref}

\begin{document}
\title{Strong correlation effects in 2D Bose-Einstein condensed dipolar
excitons}
\author{Yu.E.~Lozovik$^a$, I.L.~Kurbakov$^a$, G.E.~Astrakharchik$^b$,
J.~Boronat$^b$, and Magnus~Willander$^c$}
\affiliation{$^a$ Institute of Spectroscopy, Russian Academy of Sciences,
142190 Troitsk, Moscow region, Russia\\
$^b$ Departament de F\'isica i Enginyeria Nuclear, Campus Nord B4-B5,
Universitat Polit\`ecnica de Catalunya, E-08034 Barcelona, Spain\\
$^c$ Institute of Science and Technology (ITN), Link\"oping University,
SE-581 83, Link\"oping, T\"appan 6177, Sweden}
\date{\today}
\begin{abstract}
By doing quantum Monte Carlo {\it ab initio} simulations we show that dipolar
excitons, which are now under experimental study, actually are strongly
correlated systems. Strong correlations manifest in significant deviations of
excitation spectra from the Bogoliubov one, large Bose condensate depletion,
short-range order in the pair correlation function, and peak(s) in the
structure factor.
\end{abstract}
\maketitle
%%%%%%%%%%%%%%%%%%%%%%%%%%%%%%%%%%%%%%%%%%%%%%%%%%%%%%%%%%%%%%%%%%%%%%%%%%%%%%
Two-dimensional (2D) dipolar excitons (DEs) with spatially separated electrons
($e$) and holes ($h$) are extremely interesting due to increased lifetime,
which permits to achieve different quasi-equilibrium exciton phases predicted
for the system, e.g., 2D DE superfluid in extended systems \cite{LY,BCS,LBSF,%
VSF,BSF} and traps \cite{LKW}, Bose-Einstein condensate \cite{HD}, crystal
phase \cite{cr,MCcr} and supersolid \cite{LV}. A number of interesting
phenomena, such as Josephson effects \cite{VQD}, light backscattering and
other interesting optical properties \cite{opt} can also be observed in the
system.

DE systems can be created in coupled quantum wells (CQWs) \cite{LY} or in
a single quantum well (SQW) applying a  large external electric field (see
theory in \cite{SQW}). At present, a large experimental progress has been
achieved in the study of 2D DE collective properties in CQWs \cite{S1,S2,T,B,%
M,M1} and SQW \cite{TSQW}.

The superfluidity and coherent properties of equilibrium $e$-$h$ systems (in a
sense of "spatially separated semimetal") have been also theoretically studied
\cite{LY,BCS,VSF,VQD}. An important progress was achieved by studying an
electron bilayer in high magnetic field with $1/2+1/2$ filling of Landau
levels \cite{2DEEt}. It can be proved that the properties of the system can be
presented as a BCS state of a spatially separated (composite fermion) $e$-$h$
system at zero magnetic field \cite{LY,BCS,KK}. The predicted $e$-$h$
superfluidity and Josephson-like effects in this system were observed later on
experimentally \cite{2DEEe}. It is worth noticing that the 2D dipolar Bose
systems under consideration have been recently  realized in atomic systems
with large dipolar moments ({\it e.g.}, for Cr atoms) \cite{dipat} and polar
molecules \cite{polmol}.

The majority of theoretical models describe the excitonic Bose condensate as
an ideal or weakly correlated gas. Unfortunately, these approaches have a
very limited region of applicability. Indeed, at small densities the repulsive
dipole-dipole potential can be described by one parameter $a$, the $s$-wave
scattering length, and the properties are expected to be universal, i.e., to
be the same for all interaction potentials having the same value of scattering
length and, in particular, to be the same as in a system of hard-disks of
diameter $a$. In fact, the latter is known to be weakly correlated only in
ultra-rarified systems \cite{HD}. So, the model of weakly correlated excitons
holds only in ultra-rarified gases which have extremely low critical
temperature. For real experimental excitonic densities such models can provide
only a qualitative description. Thus, a more accurate model should be worked
out and a more precise study should be done in order to describe 2D DEs in
quantum wells (QWs).

This paper is devoted to a detailed microscopic study of 2D DEs by means of
the diffusion Monte Carlo (DMC) technique. We prove that excitons are in fact 
strongly correlated in all the main up-to-now experiments with CQW \cite{T,B,%
M} in which low-temperature collective effects in exciton luminescence have
been observed. We have obtained the following results supporting this fact:

(i) The dimensionless compressibility $\zeta=(m^3/2\pi\hbar^2)/\chi$ and
the contribution of dipole-dipole collisions to the chemical potential
$\zeta'=\mu_cm/2\pi\hbar^2n$ are much larger than unity $\zeta,\zeta'\gg1$
($\chi$ and $\mu_c$ are the corresponding dimensional quantities, $m$ is the
exciton mass, and $n$ is the total exciton density in $g_{ex}$ spin degrees;
$g_{ex}=4$ in GaAs). For weakly correlated excitons one has
$\zeta,\zeta'\ll1$.

(ii) The Bose-condensate density $n_0$ of excitons at $T=0$ is 2-4 times
smaller than the total density $n$ (on the contrary, in the weak-correlation
regime one assumes $n_0\approx n$).

(iii) There are prominent peaks in both the pair distribution function and 
the structure factor providing an evidence of short-range order. Moreover,
in the structures studied in Refs.~\cite{T,B,M} at typical exciton density,
{\it e.g.}, $n=5\cdot10^{10}$ cm$^{-2}$ two peaks in the structure factor and
three bumps of the pair distribution function are clearly visible.

(iv) The excitation spectrum strongly deviates from the weakly-correlated
Bogoliubov prediction although, even in the crossover between exciton and
$e$-$h$ regimes, there is still no roton minimum as found in
Refs.~\cite{T,B,M}.

(v) In the superfluid phase, the superfluid excitonic density $n_l$ is close
to total density $n$ \cite{nl=n}. In weakly correlated systems, the quantity
$n_l$ is  significantly smaller than $n$ at the superfluid transition.

(vi) The superfluid transition temperature in an infinite system $T_c$ is only
slightly smaller than the degeneration temperature $T_{deg}$, but the
quasi-condensation temperature is 2-2.5 times larger than $T_{deg}$
\cite{gex=4}. According to the theory of weakly correlated excitons these
temperatures are logarithmically small compared to $T_{deg}$.

(vii) The excitonic Bose condensate profile in a 2D large-size harmonic trap
at $T=0$ differs appreciably from the Thomas-Fermi inverted parabola.

For Timofeev's experiment \cite{TSQW} with SQW we find excitons to be
intermediately correlated.

We have performed DMC simulations at $T=0$ of a 2D homogeneous system with
$N=100$ DEs in a square simulation box with periodic boundary conditions and
for 12 different dimensionless densities $\bar n\equiv nx_0^2=2^{i-9}$, where
$i$ is an integer number ($1\le i\le12$), and
\begin{equation}\label{x0}
x_0=\frac{md^2}{4\pi\varepsilon\varepsilon_0\hbar^2}=
\frac{me^2D^2}{4\pi\varepsilon\varepsilon_0\hbar^2}>0
\end{equation}
is the characteristic length scale for the excitonic dipole-dipole interaction
$V(r)=\hbar^2x_0/mr^3$. Here, $d=eD>0$ is the exciton dipole moment,
$$
D=\left|\int_{-\infty}^{\infty}(|\psi_e(z)|^2-|\psi_h(z)|^2)zdz\right|
$$
is the effective $e$-$h$ separation, with $\psi_e(z)$ and $\psi_h(z)$ the
wavefunctions of $e$ and $h$ in the $Oz$ direction, $e>0$ is the hole charge,
$\varepsilon$ is the dielectric constant of the QW structure
($\varepsilon\approx12.5$ in GaAs), and $\varepsilon_0=1/4\pi$ is the
permittivity of vacuum.

Parameters of recent experiments \cite{S1,S2,T,B,M,TSQW} are summarized in
Table~\ref{table1}.

\begin{table}
\begin{small}
\begin{tabular}{llll}
\hline
structure&$D$, nm&$1/x_0^2$, cm$^{-2}$&Ref.\\
\hline
GaAs/AlGaAs CQWs&15.5&$1.56\cdot10^{10}$&\cite{S1}\\
InGaAs/GaAs CQWs&10.5&$1.83\cdot10^{11}$&\cite{S2}\\
AlAs/GaAs   CQWs&13.6&$2.64\cdot10^{10}$&\cite{T}\\
GaAs/AlGaAs CQWs&12.3&$3.94\cdot10^{10}$&\cite{B}\\
GaAs/AlGaAs CQWs&14.1&$2.28\cdot10^{10}$&\cite{M}\\
GaAs SQW&$\approx6$&$\sim7\cdot10^{11}$&\cite{TSQW}\\
\hline
\end{tabular}
\end{small}
\caption{\small Experimental structures and values of their $e$-$h$
separations $D$ and dimensional densities $n=1/x_0^2$ corresponding to
dimensionless density $\bar n=1$. The exciton mass is $m=0.14m_0$ for
InGaAs/GaAs CQWs and $m=0.22m_0$ for other CQWs and SQW, with $m_0$ being the
free electron mass).}
\label{table1}
\end{table}

We note that the excitonic collective properties are experimentally observed
in Refs. \cite{T,B,M} at dimensionless densities
$1/4\lesssim\bar n\lesssim4$. At the lowest density $\bar n=1/256$ of our
computation the superfluid crossover temperature in the DE system of Ref.~%
\cite{TSQW} is rather low, $T_c\approx0.12$ K. At the highest density we have
considered, $\bar n=8$, the dimensional density of DEs in the structure of
Ref.~\cite{S1} is rather large, $n\approx1.2\cdot10^{11}$ cm$^{-2}$.

The trial wave function used for importance sampling in DMC is similar to the
one used in our previous studies; for details of the calculation we refer the
reader to Ref.~\cite{DI}.

\begin{figure}[t]
\includegraphics[width=8.65cm,height=8.45cm]{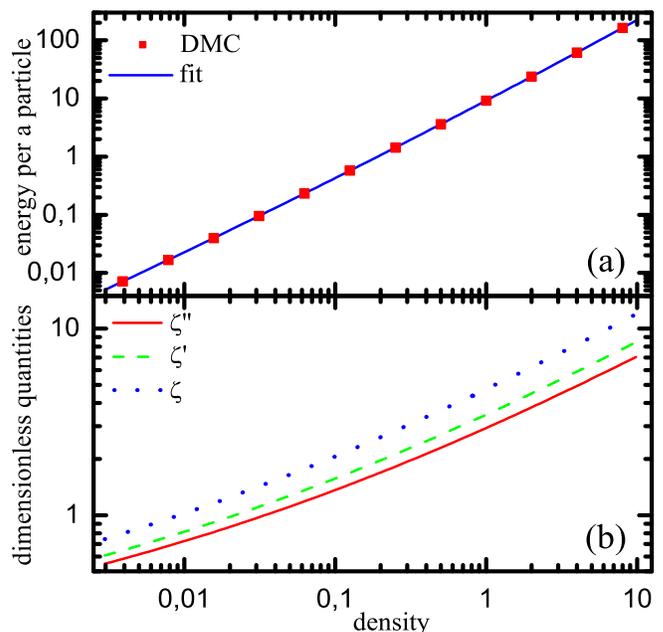}%{e.eps}
\vskip -4mm
\caption{\small (a) Ground-state energy per exciton $E_0/N$ as a function of
the density (we use $\hbar=m=x_0=1$ units; length scale $x_0$ is given by
Eq. (\ref{x0})). Squares, DMC; solid line, Eq. (\ref{Efit}). (b) Dimensionless
energy per exciton $\zeta''$ (solid line), collisional contribution to
chemical potential $\zeta'$ (dashed line) and compressibility $\zeta$ (dotted
line) calculated from Eq. (\ref{Efit}) as a function of the density.}
\end{figure}

For each density, we compute the ground state energy per exciton $E_0/N$ (see
Fig.~1(a)). The DMC results are reproduced within a 0.025\% accuracy by the
polynomial fit (now we go to $\hbar=m=x_0=1$ units, so, $n=\bar n$)
\begin{equation}\label{Efit}
E_0/N=a_e\exp(b_e\ln n+c_e\ln^2n+d_e\ln^3n+e_e\ln^4n),
\end{equation}
with $a_e=9.218$, $b_e=1.35999$, $c_e=0.011225$, $d_e=-0.00036$, and
$e_e=-0.0000281$.

Using Eq. (\ref{Efit}) we calculate the collisional contribution to the
chemical potential $\mu_c=dE_0/dN$ \cite{elst} and the inverse compressibility
$1/\chi=d\mu_c/dn$. The results for $E_0/N$, $\mu_c$ and $1/\chi$ in terms of
the quantities $\zeta''\equiv2E_0/NT_0$, $\zeta'\equiv\mu_c/T_0$ and
$\zeta\equiv n/\chi T_0$ are shown in Fig. 1(b). (Here
$T_0=g_{ex}T_{deg}=2\pi n$ is the degeneration temperature of the
spin-polarized DEs and $T_{deg}$ is the degeneration temperature of real
(spin-depolarized) DEs.) We find that at densities $1/4\lesssim n\lesssim4$
the dimensionless compressibility $\zeta$ is $3\lesssim\zeta\lesssim8$, {\it
i.e.}, $\zeta\gg1$. This is an evidence of the presence of strong correlations
between 2D CQW DEs whose collective state was investigated in Refs.~\cite{T,%
B,M}.

\begin{figure}[t]
\includegraphics[width=8.65cm,height=9.95cm]{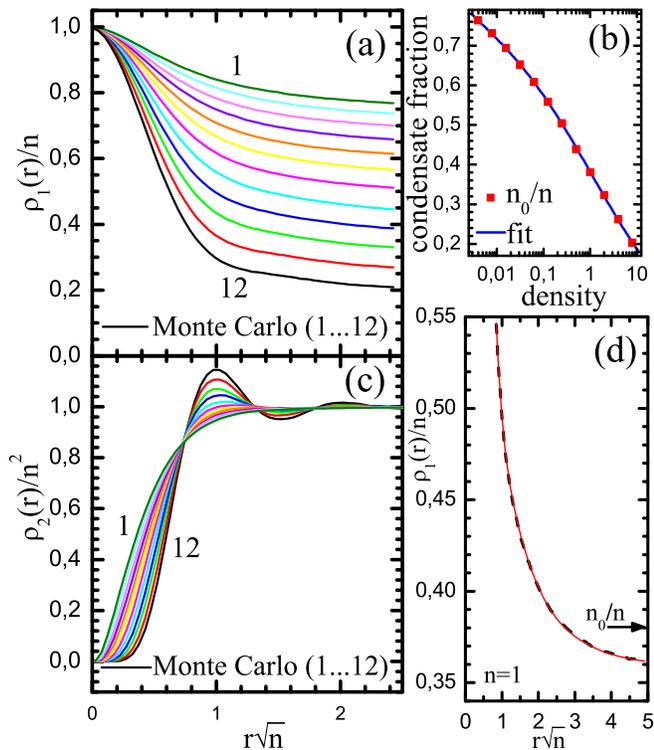}%{drpd.eps}
\vskip -4mm
\caption{\small One-body density matrix (a) and pair distribution function (c)
for the system of $N=100$ excitons at 12 densities ($n=2^{i-9}$,
$i=1,2...12$). (b) Bose-condensate fraction and polynomial fit (\ref{n0fit})
as a function of the density. (d) One-body density matrix $\rho_1({\bf r})$
shown in (a) at density $n=1$ and large distances $r$: dash line, DMC
simulation; thin solid line, hydrodynamic calculation.}
\end{figure}

In Fig.~2(a), we show results for the polar-angle-averaged one-body density
matrix
$$
\rho_1(r)=\int_0^{2\pi}\langle\hat\Psi^+({\bf r})\hat\Psi(0)\rangle
d\varphi/2\pi.
$$
Here $\hat\Psi({\bf r})$ is the exciton field operator and the brackets
$\langle...\rangle$ denote ground-state averaging.

The asymptotic behavior of $\rho_1(r)$ matches the result derived from a
hydrodynamic calculation (see Fig.~2(d)) giving us additional confidence on
the accuracy of the DMC simulation.

In Fig.~2(b) we plot DMC results of the Bose-condensate fraction $n_0/n$ and
the corresponding polynomial fit on top of the data, also in terms of $\ln n$,
\begin{equation}\label{n0fit}
n_0/n=a_0^n\exp(b_0^n\ln n+c_0^n\ln^2n+d_0^n\ln^3n),
\end{equation}
with parameters $a_0^n=0.3822$, $b_0^n=-0.2342$, $c_0^n=-0.02852$, and
$d_0^n=-0.001594$. The results reveal a strong Bose condensate depletion at
densities $1/4\lesssim n\lesssim4$. This is again a clear manifestation of
strong correlations between 2D CQW DEs.

In Fig.~2(c), we show the results of DMC simulations for polar-angle-averaged
excitonic pair distribution function
$$
\rho_2(r)=\int_0^{2\pi}\langle\hat\rho({\bf r})\hat\rho(0)\rangle
d\varphi/2\pi
$$
($\hat\rho({\bf r})=\hat\Psi^+({\bf r})\hat\Psi({\bf r})$ is the excitonic
density operator). The prominent bump in the pair distribution function at
densities $1/4\lesssim n\lesssim4$ is an evidence of both the short-range
order and the strong correlations present in DE systems.

\begin{figure}[t]
\includegraphics[width=8.65cm,height=9.2cm]{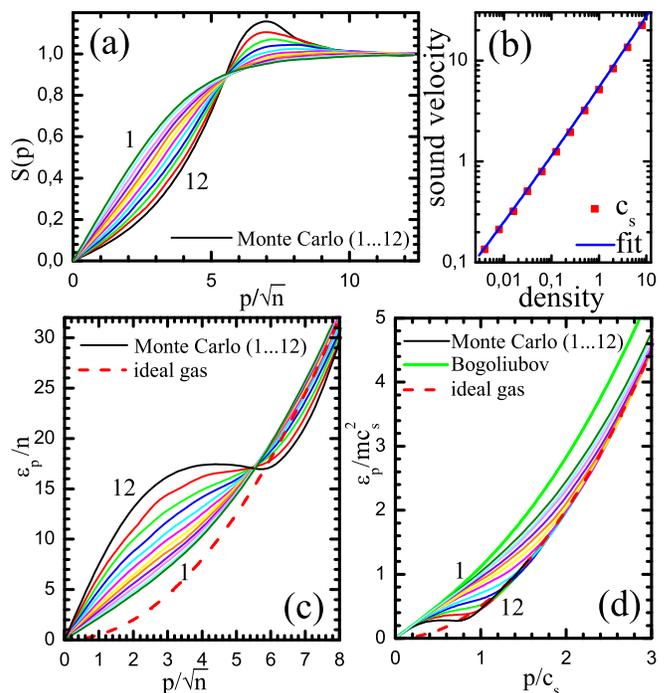}%{spep.eps}
\vskip -4mm
\caption{\small Exciton structure factor (a) and excitation spectrum (c)
for 12 densities ($n=2^{i-9}$, $i=1,2...12$). (b) Sound velocity calculated
from the slope of the structure factor (squares) and from Eq. (\ref{Efit})
(solid line) as a function of the density. (d) Comparison of excitation
spectra at different densities ($n=2^{i-9}$, $i=1,2...12$) with the Bogoliubov
prediction.}
\end{figure}

The structure factor $S(p)$ is related to the Fourier transform of the pair
distribution function $\rho_2(r)$. We plot the DMC results of $S(p)$ in
Fig.~3(a). A sharp peak in $S(p)$ at densities $1/4\lesssim n\lesssim4$ is 
consequence of the short-range order in 2D CQW DE systems.

Using the DMC results of the structure factor $S(p)$ one can predict an upper
bound for the excitation spectrum by using the Feynman formula \cite{FF}
\begin{equation}\label{ep}
\varepsilon_p\le\frac{p^2}{2S(p)}.
\end{equation}
(see Fig.~3(c)). We see that the collective excitation spectrum of the 2D DE
system is rather far from the Bogoliubov form (see Fig.~3(d)) in the range
$1/4\lesssim n\lesssim4$. This fact suggests the relevant role of correlations
in the excitonic collective state in CQWs.

We compare two calculations of sound velocity $c_s$: from the low-momentum
structure factor slope ($c_s=p/2S(p\to0)$; see Eq. (\ref{ep})) and from the
compressibility $1/\chi$ ($c_s=\sqrt{n/\chi}$), where $\chi$ is calculated by
differentiation of the energy fit (\ref{Efit}). The good agreement between the
two calculations (see Fig.~3(b)) is expected from an exact calculation like
the present one and supposes a crucial internal consistency check.

\begin{figure}[t]
\includegraphics[width=8.65cm,height=8.75cm]{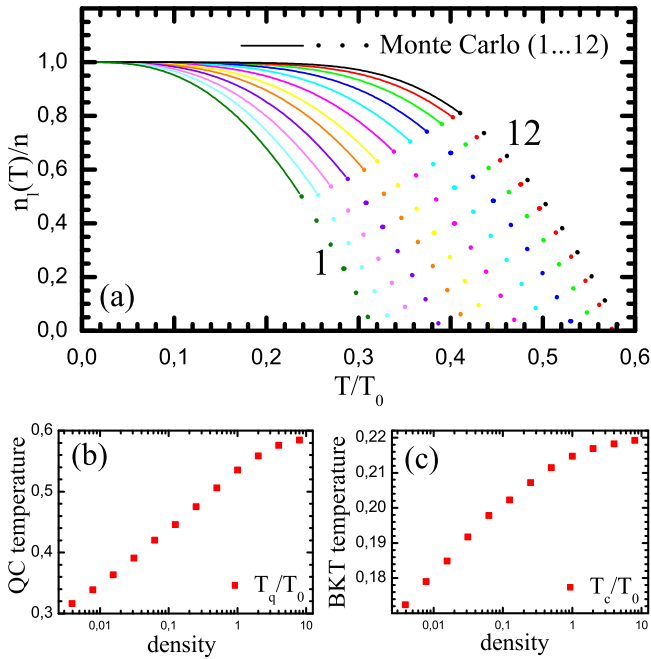}%{nl.eps}
\vskip -4mm
\caption{\small (a) Temperature dependence of the local excitonic superfluid
fraction in $g_{ex}=4$ spin degrees at 12 densities ($n=2^{i-9}$,
$i=1,2...12$; solids). Dotted lines depict the formal extrapolation of the
calculation to the quasi-condensation crossover region. (b) Quasi-condensation
temperature and (c) BKT transition temperature in the infinite system as a
function of the density.}
\end{figure}

In Fig.~4(a), we show results derived from the Landau formula of the
temperature dependence of the {\it local} (vortex-unrenormalized \cite{VR,%
jc061181,GS}) superfluid fraction of 2D quasi-condensed \cite{Popov} DEs
$n_l(T)/n$ far from the quasi-condensation crossover. Results for the
Berezinskii-Kosterlitz-Thouless (BKT) superfluid transition temperature
\cite{Berezinskii,jc061181,NKT} in a 2D infinite system $T_c$ as well as the
quasi-condensation temperature $T_q$ are shown in Figs.~4(b,c).

We see that at densities $1/4\lesssim n\lesssim4$ the local superfluid
fraction $n_l/n$ at $T\le T_c$ is close to unity. Moreover, the
quasi-condensation temperature $T_q$ at $1/4\lesssim n\lesssim4$ is 2-2.5
times higher than the degeneration temperature $T_{deg}=T_0/g_{ex}=0.25T_0$
and the BKT superfluid transition temperature $T_c\approx0.22T_0$ is only
slightly smaller than $T_{deg}$. All these features are manifestations of the
strong correlations between 2D CQW DEs. Besides, our results show that {\it
in real, strongly correlated, DE systems quasi-condensate and superfluid
phases are much easier to achieve experimentally than if DEs were weakly
correlated}. (For weakly correlated DEs temperatures $T_c$ and $T_q$ are
logarithmically small in comparison with $T_{deg}$ \cite{Popov,b3704936}).

\begin{figure}[t]
\includegraphics[width=8.65cm,height=9.1cm]{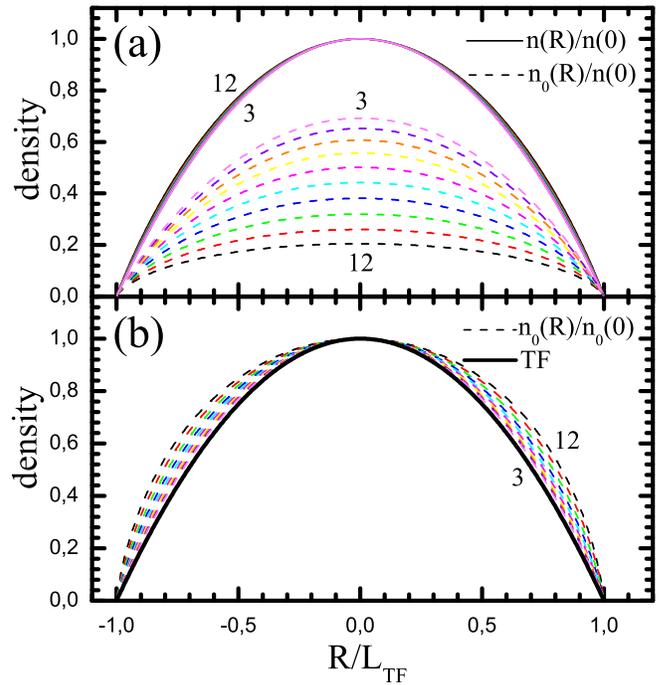}%{harm.eps}
\vskip -4mm
\caption{\small (a) Total and Bose condensate profiles of harmonically trapped
excitons in LDA at $\zeta'_e=10$ for 10 densities in the trap center
($n(0)=2^{i-9}$, $i=3,4...12$). (b) Comparison of the Bose-condensate
profiles at different $n(0)$ with the TF inverted parabola.}
\end{figure}

Based on our DMC simulations of homogeneous infinite systems (Eqs.
(\ref{Efit}) and (\ref{n0fit})) and applying local density approximation
(LDA) we have calculated the total $n({\bf R})$ and Bose-condensate
$n_0({\bf R})$ exciton profiles in a harmonic trap of large size
($\sqrt{n(0)L_{TF}^2}\gg1$) at $T=0$. Here, $L_{TF}=\sqrt{2N/\pi n(0)}$ is
Thomas-Fermi (TF) radius and $N\gg1$ is number of DEs in the trap. The results
obtained for both functions are shown in Fig.~5(a).

If the electrostatic contribution is large
$\zeta'_e\equiv\mu_e/T_0=2e^2D/\varepsilon=10$ \cite{elst} (this is its
typical value for the experiments of Refs.~\cite{T,B,M}), the total profile is
in very good agreement with the TF inverted parabola (see Fig. 5(a))
\cite{elstTF}. However, the Bose condensate profile at
$1/4\lesssim n\lesssim4$ differs essentially from the inverted parabola (see
Fig.~5(b)). The latter fact shows a failure of the theory of weakly correlated
excitons for which the difference between Bose condensate profile and TF
inverted parabola is logarithmically small \cite{HD}.

Correlation effects in a dense 2D DE gas in CQWs can be detected by
observation of the exciton luminescence in in-plane magnetic field. Indeed, in
absence of magnetic field, the momentum of a recombined exciton $p$ according
to the momentum conservation in one-photon exciton recombination is equal to
the QW plane projection [$(\hbar\omega/c_0)\sin\theta$] of the momentum of the
emitted photon $\hbar\omega/c_0$ \cite{h=m=x0=1}. Here $\theta$ is the angle
between the normal and emitted photon in {\it free space}, $c_0$ is the light
velocity in free space, and $\omega$ is the photon frequency. However, if
there is in-plane magnetic field $H_{\parallel}$, the dispersion curve
$\varepsilon_p$ {\it is shifted} by the quantity $p_H=eDH_{\parallel}/c_0$
\cite{BH}. This results in the following connection between the photon angle
$\theta$ and the exciton momentum ${\bf p}$:
\begin{equation}\label{ppH}
(\hbar\omega/c_0)\sin\theta=|{\bf p}-{\bf p}_H|.
\end{equation}
If one considers only normal luminescence ($\theta=0$), then $p=p_H$ 
(\ref{ppH}), and thus $\varepsilon_p=\varepsilon_{p_H}$. Hence, for
spectral-angle luminescence along the normal in the field $H_{\parallel}$ one
obtains
\begin{equation}\label{It0o}
I_{\theta=0}^H(\omega)=I_{\theta=0}^H
\delta(\omega-\Omega+\varepsilon_{p_H}/\hbar)\;\;\;\;(p_H\gg p_T),
\end{equation}
where $I_{\theta=0}^H$ is the spectrally integrated angle luminescence along
the normal in the field $H_{\parallel}$, $\Omega$ is the exciton resonance
frequency for luminescence at $H_{\parallel}=\theta=0$, and $p_T=T/c_s$ is
the typical thermal momentum of Bose condensed DEs which is the boundary
between thermal $p\ll p_T$ and zero-temperature $p_T\ll p\ll\hbar\sqrt n$
regions of quasi-condensate phase fluctuations \cite{Popov}.

From Eq. (\ref{It0o}), we find that the observation of the
magnetic-field-induced normal excitonic luminescence line shift enables one to
{\it directly measure the excitation spectrum} $\varepsilon_p$. In the
structure of Ref.~\cite{BH} ($D=12.3$ nm) the density of the exciton gas is
$n=2\cdot10^{10}$ cm${}^{-2}$, the field range $0\le H_{\parallel}\le8$ T
corresponds to the momentum range $0\le p_H/\hbar\le1.5\cdot10^6$ cm${}^{-1}$.
This range covers both long-wave-length (hydrodynamic) and intermediate scales
up to short-wave-length (ideal-gas) ones on which
$\varepsilon_{p_H}\approx p_H^2/2m$. Note that the thermal momentum $p_T$ at
$T=0.2T_0=1$ K $<T_c,T_q$, $m=0.22m_0$ and $\zeta=3.7$ (see above) corresponds
to a field $H_{\parallel}^T=(c_0T/eD)\sqrt{m/\zeta T_0}\approx0.2$ T.

By measuring the slope of the excitation spectrum $\varepsilon_{p_H}$ one can
obtain sound velocity $c_s\approx\varepsilon_{p_H}/p_H$ ($p_H\ll\hbar\sqrt n$;
see Fig.~3(c)). The dimensionless exciton compressibility is then found as
$$
\zeta=m^2c_s^2/2\pi\hbar^2n.
$$
Alternatively, one can determine the dimensionless compressibility $\zeta$ by
measuring the magnetic field dependence of spectrally integrated angle
luminescence along the normal
\begin{equation}\label{IpHz}
I_{\theta=0}^H=I_0\frac{q_r^2c_0}{4\pi eD\sqrt{mT_0}}
\frac{\sqrt{\zeta}}{H_{\parallel}}\;\;\;\;
(p_T\ll eDH_{\parallel}/c_0\ll\hbar\sqrt n).
\end{equation}
Here $q_r=\hbar\omega/c_0\approx\hbar\Omega/c_0$ is the radiation zone in free
space, $I_0=\kappa(\hbar\Omega/\tau_0)n_0$ is the Bose-condensate luminescence
at $T=0$, $\tau_0$ is the lifetime of a spin-depolarized exciton at $T=0$ in
pure CQW structure, and $\kappa=1/2$ ($\kappa=1$) if the luminescence is
measured on one (both) sides of CQW plane.

At $n=2\cdot10^{10}$ cm${}^{-2}$, $D=12.3$ nm, $T=0.1T_0=0.5$ K, $m=0.22m_0$
and $\zeta=3.7$ the momentum range $p_T\ll eDH_{\parallel}/c_0\ll\hbar\sqrt n$
in (\ref{IpHz}) corresponds to the field range
$(c_0T/eD)\sqrt{m/\zeta T_0}\ll H_{\parallel}\ll c_0\hbar\sqrt n/eD$, or, 0.1
T $\ll H_{\parallel}\ll$ 0.75 T.

The Bose condensate luminescence at $T=0$ entering into Eq. (\ref{IpHz}) can
be approximately calculated as thermal-phase-fluctuation quasi-condensate
luminescence at low temperatures
\begin{equation}\label{I0}
I_0\approx\int_0^{p_T}I_{\theta=0}^H\frac{2\pi p_Hdp_H}{q_r^2}=
\kappa\frac{\hbar\Omega}{\tau_0}n_0\;\;\;\;(T\ll T_q),
\end{equation}

Finally, the measurement of the quantity (\ref{I0}) enables one to determine
an important parameter of the exciton interaction, the Bose-condensate density
at $T=0$
\begin{equation}
n_0=\frac1{\kappa}\frac{\tau_0}{\hbar\Omega}I_0.
\end{equation}

In conclusion, Bose-Einstein condensation, superfluidity, and microscopical
properties of 2D dipolar excitons in SQW and CQWs have been studied by {\it ab
initio} simulations and analytical calculations. In all the recent experiments
on excitonic Bose condensation in CQWs the excitons are strongly correlated.
We have shown that in a strongly correlated regime the excitonic collective
state is much easier to be achieved than in weakly correlated DEs. Our
calculation of the microscopical properties of 2D DEs enables one to
investigate quantitatively QW excitonic Bose-Einstein condensation,
superfluidity, luminescence and nonlinear optical effects by means of the
hydrodynamic method in quantum field theory.

\section*{Acknowledgements}
The work was supported by the Russian Foundation of Basic Research, Swedish
Foundation for Strategic Research (SSR), by (Spain) Grant No. FIS2005-04181,
Generalitat de Catalunya Grant No. 2005SGR-00779 and RFBR. G.E.A. acknowledges
post doctoral fellowship by MEC (Spain).

%%%%%%%%%%%%%%%%%%%%%%%%%%%%%%%%%%%%%%%%%%%%%%%%%%%%%%%%%%%%%%%%%%%%%%%%%%%%%%

\begin{thebibliography}{99}
\bibitem{nl=n}
In 2D this property is specific for strongly (but not too strongly) and
intermediately correlated Bose systems.
\bibitem{elst}
The chemical potential has two contributions $\mu=\mu_c+\mu_e$. One is of
electrostatic nature $\mu_e=4\pi ne^2D/\varepsilon$ and associated with the
energy of a parallel-sided capacitor forming by charged $e$ and $h$ layers
which are separated by the distance $D$. The other contribution $\mu_c$ is due
to in-plane dipole-dipole collisions.
\bibitem{GS}
True (vortex-renormalized) superfluid fraction is calculated in
Ref.~\cite{LKW}a.
\bibitem{elstTF}
In the homogeneous system the large value of the electrostatic contribution
$\mu_e$ results in an almost linear density dependence of the total chemical
potential. This suppresses the nonlinear density dependence due to in-plane
collisions $\mu_c$ and results in an almost linear overall dependence. So, the
total density profile in the harmonic trap is close to TF inverted parabola.
\bibitem{h=m=x0=1}
In the following we will use usual (dimensional) units.
\end{thebibliography}
\end{document}